\documentstyle[prl,aps, epsfig]{revtex}
\begin{document}
\newcounter{TC}
\newcommand{\useTC}[1]{\refstepcounter{TC}\label{#1}\arabic{TC}}
\newcommand{\MeVcsq}{${\rm MeV}/c^2$}
\newcommand{\GeVc}{${\rm GeV}/c$}
\newcommand{\EGEV}{${\rm GeV}/c$}
\newcommand{\PGEV}{${\rm GeV}^2/c^2$}
\newcommand{\MGEV}{${\rm GeV}/c^2$}

\draft
\title{Observation of exotic meson production in the reaction
$ \pi^{-} p \rightarrow \eta^{\prime} \pi^- p$ at 18 \GeVc}
\author{
E.~I.~Ivanov\rlap,$^1$ D.~L.~Stienike\rlap,$^1$
D.~I.~Ryabchikov\rlap,$^3$ G.~S.~Adams\rlap,$^7$
T.~Adams\rlap,$^{1}$  
Z.~Bar-Yam\rlap,$^4$ J.~M.~Bishop\rlap,$^1$
V.~A.~Bodyagin\rlap,$^5$
D.~S.~Brown\rlap,$^{6}$ 
N.~M.~Cason\rlap,$^1$
S.~U.~Chung\rlap,$^2$
J.~P.~Cummings\rlap,$^{7}$ K.~Danyo\rlap,$^2$
S.~P.~Denisov\rlap,$^3$ V.~A.~Dorofeev\rlap,$^3$
J.~P.~Dowd\rlap,$^4$
P.~Eugenio\rlap,$^{4}$ 
X.~L.~Fan\rlap,$^6$
R.~W.~Hackenburg\rlap,$^2$ M.~Hayek\rlap,$^4$  D. Joffe\rlap,$^6$
 I.~A.~Kachaev\rlap,$^3$
 W.~Kern\rlap,$^4$
E.~King\rlap,$^4$ O.~L.~Kodolova\rlap,$^5$
V.~L.~Korotkikh\rlap,$^5$ M.~A.~Kostin\rlap,$^5$ J.~Kuhn\rlap,$^7$
V.~V.~Lipaev\rlap,$^3$ J~.M.~LoSecco\rlap,$^1$
J.~J.~Manak\rlap,$^1$ J.~Napolitano\rlap,$^7$ M.~Nozar\rlap,$^7$
C.~Olchanski\rlap,$^2$ A.~I.~Ostrovidov\rlap,$^5$
T.~K.~Pedlar\rlap,$^6$ A.~V.~Popov\rlap,$^3$
L.~I.~Sarycheva\rlap,$^5$
K.~K.~Seth\rlap,$^6$  X.~Shen\rlap,$^6$
N.~Shenhav\rlap,$^{4}$ 
W.~D.~Shephard\rlap,$^1$ N.~B.~Sinev\rlap,$^5$
J.~A.~Smith\rlap,$^7$
S.~A.~Taegar\rlap,$^{1}$ 
A.~Tomaradze\rlap,$^6$
I.~N.~Vardanyan\rlap,$^5$
D.~P.~Weygand\rlap,$^{8}$ 
D.~B.~White\rlap,$^7$ H.~J.~Willutzki\rlap,$^2$ 
M.~Witkowski\rlap,$^7$ A.~A.~Yershov\rlap,$^5$ 
}
\author{The E852 Collaboration}
\address{$^1$University of Notre Dame, Notre Dame, IN 46556, USA\\
$^2$Brookhaven National Laboratory, Upton, Long Island, NY 11973,
USA\\
$^3$Institute for High Energy Physics, Protvino, Russian
Federation\\ $^4$University of Massachusetts Dartmouth, North
Dartmouth, MA 02747, USA\\ $^5$Moscow State University, Moscow,
Russian Federation\\ $^6$Northwestern University, Evanston, IL
60208, USA\\ $^7$Rensselaer Polytechnic Institute, Troy, NY 12180,
USA \\ $^8$Thomas Jefferson National Accelerator Facility, Newport
News, VA 23606, USA}
\date{\today}

\maketitle
\vskip 1cm
\begin{abstract}
An amplitude analysis of an exclusive sample of 5765 events from
the reaction $\pi^{-} p \rightarrow \eta^{\prime} \pi^- p$~ at 18
\GeVc\ is described. The $\eta^{\prime} \pi^-$ production is
dominated by natural parity exchange and by three partial waves:
those with $J^{PC} = 1^{-+}, 2^{++},$~and $4^{++}$. A
mass-dependent analysis of the partial-wave amplitudes indicates
the production of the $a_2(1320)$ meson as well as the $a_4(2040)$
meson, observed for the first time decaying to
$\eta^{\prime}\pi^-$.  The dominant, exotic (non-$q\overline{q}) $
$1^{-+}$ partial wave is shown to be resonant with a mass of
$1.597 \pm 0.010^{+0.045}_{-0.010}$ \MGEV\ and a width of $0.340
\pm 0.040 \pm 0.050$ \MGEV\ . This exotic state, the
$\pi_1(1600)$, is produced with a $t$ dependence which is
different from that of the $a_2(1320)$ meson, indicating
differences between the production mechanisms for the two states.
\end{abstract}
 \twocolumn
Exotic mesons - those whose valence structure is not composed of a
quark-antiquark  ($q\overline{q}) $ pair -- have been
discussed\cite{jaffe,baclose,th:barnes,th:ip,qcdsum,th:bcs,th:clp,th:diquark,th:lgt,barnesmeson2000}
for many years but have only recently been observed
experimentally. The underlying structure of the observed exotic
states at 1.4 \MGEV\ decaying into
$\eta\pi^-$\cite{thompson,long_paper,cbarrel} and at 1.6 \MGEV\
decaying into $\rho^0\pi^-$\cite{sasha} is not yet understood.
Possible explanations for these $I=1$ states could be that they
are hybrid mesons, consisting of a $q\overline{q} $ pair and a
constituent gluon, or four-quark ($q\overline{q}q\overline{q}) $
states. However, within the framework of the flux-tube model the
masses of these states are somewhat low to be hybrid
mesons\cite{th:bcs}; and four-quark states are expected to be very
broad\cite{jaffe}.

Since the models for exotic mesons typically predict masses,
widths, and branching ratios, and since it is important to
classify the exotic states  to provide necessary input to the QCD
models, there is a strong motivation to search for additional
states as well as to search for additional decay modes for the
observed states. In this paper, we describe the search for exotic
states decaying into the $\eta^{\prime} \pi^-$ final state using
the reaction $\pi^{-} p \rightarrow \eta^{\prime} \pi^- p$, where
$\eta^{\prime} \rightarrow \eta \pi^{+}\pi^{-}$ and $\eta
\rightarrow \gamma \gamma$.

The data sample was collected during the 1995 run of experiment
E852 at the Multi-Particle Spectrometer facility at Brookhaven
National Laboratory (BNL).  A $\pi^{-}$ beam with laboratory
momentum 18 \GeVc\ and a liquid hydrogen target were used.  A
detailed description of the experimental apparatus can be found
elsewhere\cite{long_paper}.

\begin{center}
\begin{figure}
    \psfig{figure=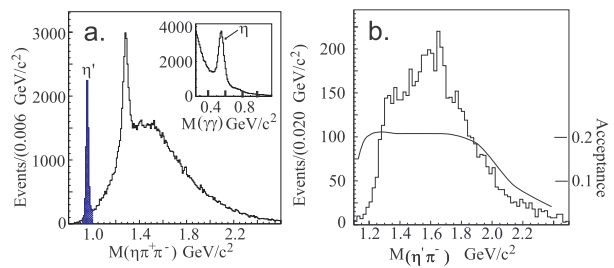,width=8.6cm,angle=0}
    \caption{(a) The $\eta\pi^+\pi^-$ effective mass distribution for
events consistent with the reaction $\pi^- p \rightarrow p \eta
\pi^+ \pi^+ \pi^- $ (two entries per event).  The inset shows the
$\gamma\gamma$ effective mass distribution in 0.01 \MGEV\ bins.
(b) The $\eta^\prime \pi^-$ effective mass distribution.  The
distributions are uncorrected for acceptance.  The smooth curve in
(b) shows the true mass acceptance based upon the angular
distributions determined in the partial wave analysis.
\label{signal}}
\end{figure}
\end{center}

The trigger required three forward-going charged tracks, a charged
recoil track and a signal in a lead-glass electromagnetic
calorimeter (LGD).  A total of 165 million triggers of this type
were recorded.  After reconstruction, 1.37 million events
satisfied the trigger topology and had two clusters in the LGD.
The $\eta$ signal is seen in the $\gamma \gamma$ effective mass
distribution in Fig.\ref{signal} (a). Applying kinematic fitting
\cite{squaw}, some 70000 events
consistent with the $p\eta \pi^{+}\pi^{-}\pi^{-}$
$(\eta\rightarrow \gamma \gamma)$ final state were found.  These
events satisfied energy-momentum conservation at the production
and $\eta$ decay vertices with a confidence level $CL>0.05$ as
well as the requirement that the difference between the azimuthal
angles of the the fitted proton direction and the measured proton
track
be less than $10^{o}$.  As seen from the $\eta \pi^{+}\pi^{-}$
effective mass distribution (uncorrected for acceptance) in
Fig.\ref{signal} (a) the $\eta^{\prime}$ signal lies over an
approximately 10\% non-$\eta^{\prime}$ background.  The second
peak in the $\eta \pi^{+}\pi^{-}$ mass spectrum is due to
production of the $f_{1}(1285)$ and $\eta(1295)$ resonances.

The next level of selection identified 6040 events consistent with
the $p\eta^{\prime}\pi^{-}$  $(\eta^{\prime}\rightarrow \eta
\pi^{+}\pi^{-}, \;\eta\rightarrow \gamma \gamma)$ final state.
These events satisfy energy-momentum conservation at the
production, $\eta^{\prime}$ and $\eta$ decay vertices with
$CL>0.05$ as well as topological and fiducial volume cuts. The
resulting uncorrected  $\eta^{\prime}\pi^{-}$ mass spectrum
(Fig.\ref{signal} (b)) has a broad peak near 1.6 \MGEV\ and
structure around 1.3 \MGEV.

\begin{center}
\begin{figure}
    \psfig{figure=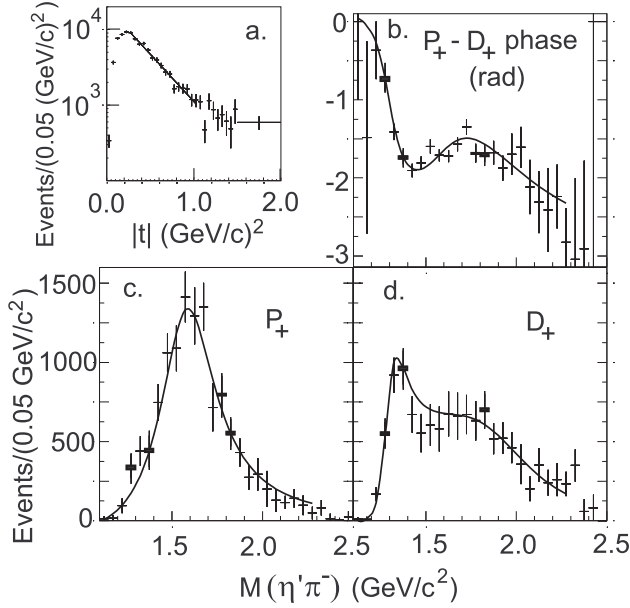,width=8.6cm,angle=0}
    \caption{(a) The acceptance-corrected $|t|$ distribution fitted with
the function $f(t)=ae^{b|t|}$ (solid line).  (b), (c), (d) The
results of the mass-independent PWA (horizontal lines with error
bars) and a typical mass-dependent fit (solid curve) using 0.05
\MGEV\ mass bins.  Only $P_+$ and $D_+$ partial waves and their
phase difference are shown.  The range of the ambiguous solutions
is plotted with black rectangles.  (b) The $(P_{+}-D_{+})$ phase
difference.  (c) The intensity distribution of the $P_+$ partial
wave. (d) The intensity distribution of the $D_+$ partial wave.
The solid curves in (b), (c), (d) show a mass-dependent fit (Fit
1) to the $P_{+}$ and $D_{+}$ wave intensities and the
$(P_{+}-D_{+})$ phase difference.
\label{t_and_pwa_50}}
\end{figure}
\end{center}

The acceptance-corrected distribution of the
four-momentum-transfer $|t|$ is shown in Fig.\ref{t_and_pwa_50} (a).
The amplitude analysis discussed below was made for the data in
the range $0.09< |t| <2.5$ \PGEV.  Because of the very low
acceptance in the region $|t| <0.09$ \PGEV, the 275 events in that
region were not used.  In the interval $0.25< |t| <1.0$ \PGEV\ the
$|t|$ distribution has an exponential behavior and can be fitted
with the function $f(t)=ae^{b|t|}$ with $b=-2.93\pm 0.11~
(\rm{GeV}/c)^{-2}$.
The  magnitude of $b$ is significantly less than that observed for
the $\eta\pi^-$ final state~\cite{thompson,long_paper}, where
$b\approx -5~ (\rm{GeV}/c)^{-2}$ (see the discussion below).


A mass-independent partial-wave analysis (PWA)
\cite{long_paper,newbnl,th:SUtwo} of the data was used to study
the spin-parity structure of the $\eta^{\prime}\pi^{-}$ system.
The partial waves are parameterized by a set of five numbers:
$J^{PC} m^{\epsilon}$, where $J$ is the angular momentum, $P$ the
parity and $C$ the C parity of the $\eta^{\prime}\pi^{-}$ system;
$m$ is the absolute value of the angular momentum projection; and
$\epsilon$ is the reflectivity (coinciding with the naturality of
the exchanged particle\cite{chungtru}).  We will use simplified
notation in which each partial wave is denoted by a letter,
indicating the $\eta^{\prime}\pi^{-}$ system's angular momentum in
standard spectroscopic notation, and a subscript, which can take
the values $0,\;+$, or $-$,  for $m^{\epsilon}=0^{-}$, $1^{+}$, or
$1^{-}$ respectively.
We assume that the
contribution from partial waves with $m> 1$ is small and can be
neglected.

Mass-independent PWA fits shown in this paper are carried out in
0.05 and 0.10 \MGEV\ mass bins from  1.1 to 2.5 \MGEV\ and all use
the $S_0,\;P_-,\;P_0,\;P_+,\;D_-,\;D_0,\;D_+$ and $G_+$ partial
waves. For each partial wave the complex production amplitudes are
determined from an extended maximum likelihood fit\cite{th:SUtwo}.
The spin-flip and spin-nonflip contributions to the baryon vertex
lead to a production spin-density matrix with maximal rank two. A
rank two mass-independent PWA in a system of two pseudoscalars
cannot be performed because of the presence of a continuous
mathematical ambiguity.  Rank-two fits were done when additional
assumptions for the amplitudes were introduced (assumptions
regarding the t-dependence and mass dependence of the amplitudes)
to resolve the continuous ambiguity problem,
and they gave results consistent with those from the rank one
fits.
The PWA fits presented in this paper
are with spin-density matrix of rank one.

The experimental acceptance was determined by comparison of the
data with a Monte Carlo event sample.  The Monte Carlo events were
generated with isotropic angular distributions in the
Gottfried-Jackson frame.  The detector simulation was based on the
E852 detector simulation package SAGEN\cite{thompson,long_paper}.
The experimental acceptance was incorporated into the PWA by means
of Monte Carlo normalization integrals\cite{long_paper}. The
quality of the fits was determined by a $\chi^2$ comparison of the
experimental multipole moments with those predicted by the results
of the PWA fit\cite{my_thesis}.


%

\begin{center}
\begin{figure}
    \psfig{figure=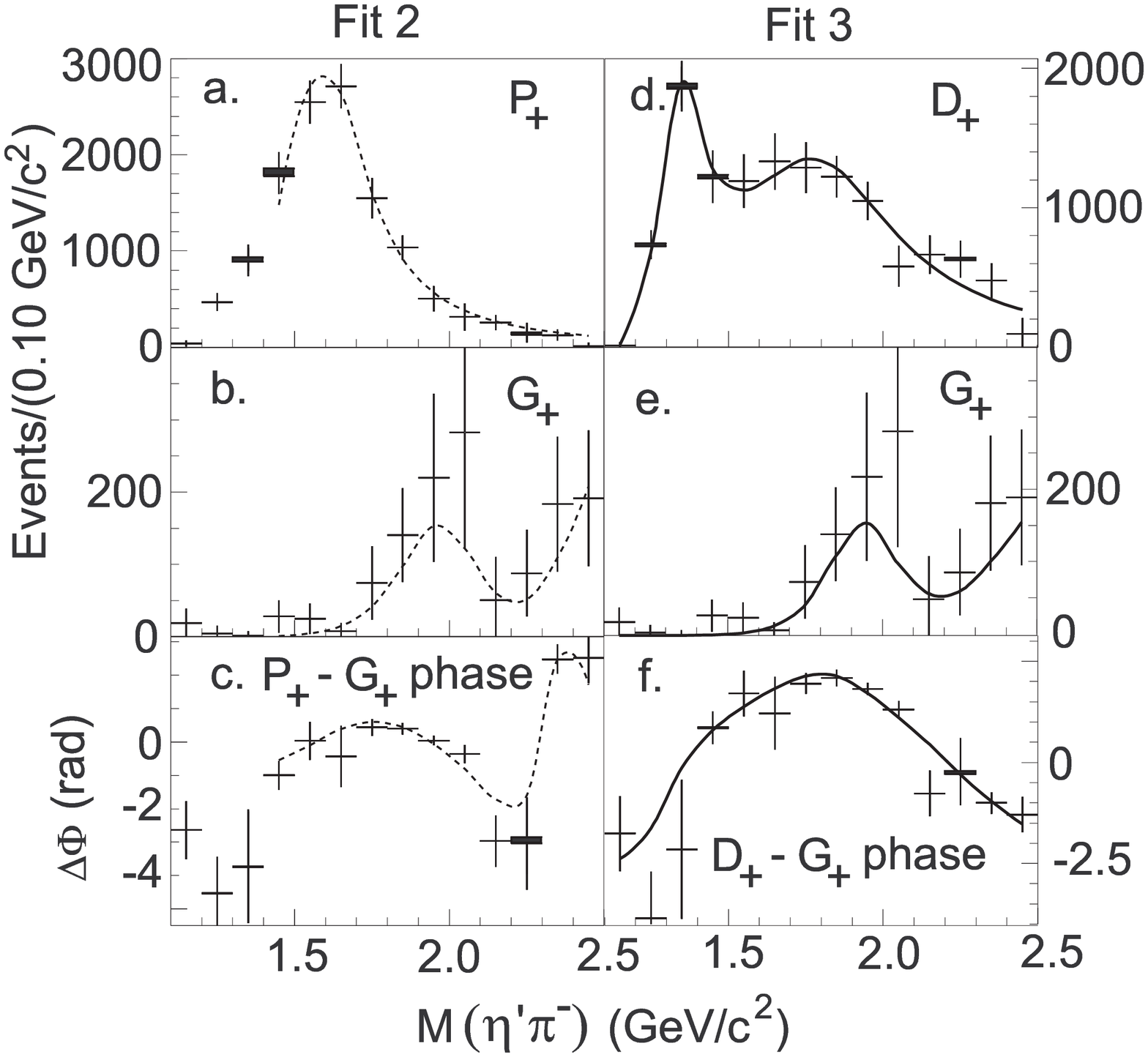,width=8.6cm,angle=0}
    \caption{ The results of the mass-independent PWA (horizontal
lines with error bars) and typical mass-dependent fits (solid and
dashed curves) using 0.1 \MGEV\ mass bins. (a)-(c) show the PWA
and Fit 2 (dashed curves) for the $P_+$ - $G_+$ intensities and
their phase difference;  (d)-(f) show the PWA and Fit 3 (solid
curves) for the $D_+$ - $G_+$ intensities and their phase
difference. The range of the ambiguous solutions is plotted with
black rectangles.
%
\label{mda_one}}
\end{figure}
\end{center}

Results of the PWA are shown in Fig.~\ref{t_and_pwa_50} for the
0.05 \MGEV\ fits and Fig.~\ref{mda_one} for the 0.10 \MGEV\ fits.
The former are intended to show detail in the high statistics
low-mass region and the latter are used to study the high-mass
region. The unnatural-parity-exchange waves (not shown) are small,
poorly determined, and do not affect our results.

The acceptance-corrected numbers of events predicted by the PWA
fits for the stronger partial waves and the phase differences
between them are shown as points with error bars.
There are discrete mathematical ambiguities in the description of
a system of two pseudoscalar mesons\cite{th:SUtwo}. The ambiguous
solutions were found by performing 1000 PWA fits with random
starting values in each mass bin. The range of the ambiguous
solutions in a mass bin is presented by black rectangles, and the
maximum extent of their statistical errors is shown as the error
bar.  In most mass bins, the range of values found for the
ambiguous solutions was small enough that they cannot be
distinguished in the figures.

Between 1.5 and 1.8 \MGEV\ the exotic $P_+$ is the dominant wave.
Its intensity distribution consists of a broad structure peaked
near 1.6 \MGEV\ .  The $D_+$ wave intensity has a narrow peak at
1.3 \MGEV\ associated with the $a_2(1320)$ and a broad structure
at higher masses.  The $G_+$ wave intensity is negligible   below
1.7 \MGEV\ and is clearly nonzero in the higher mass region. The
$(P_+-D_+),\;\;(D_+-G_+)$ and $(P_+-G_+)$ wave phase differences
show rapid changes possibly indicative of the presence of
interfering resonant states.  The observed $P_+$ and $D_+$
intensities and their relative phase difference distribution are
consistent with those reported by the VES\cite{ex:VESetapm}
collaboration. Leakage studies\cite{long_paper} were carried out
and no leakage of significance was found among the dominant waves.

To study the nature of the observed partial waves, three different
kinds of mass-dependent analyses (MDA)\cite{long_paper} have been
carried out. In the first fit type (Fit 1),  the $P_+$ and $D_+$
intensities and their phase difference were fitted using the PWA
results in 0.05 \MGEV\ bins.  For the other two types (Fit 2 and
Fit 3), the $P_+$ and $G_+$, and the $D_+$ and $G_+$ waves
respectively were fitted, along with their phase differences,
using the 0.10 \MGEV\ PWA results. These fits all used linear
combinations of relativistic Breit-Wigner functions (poles) with
mass-dependent widths and Blatt-Weisskopf barrier factors.

Because of the presence of distinct ambiguous PWA solutions in
some mass bins, all possible combinations of these solutions were
used as inputs to the MDA fits. Typical fits are shown as the
smooth curves in Figs.~\ref{t_and_pwa_50} and \ref{mda_one}.
The fits use a single Breit-Wigner function to describe the $P_+$
wave, and two Breit-Wigner functions to describe both the $D_+$
and $G_+$ partial waves.  (The fits can be improved by a second
$P_+$ resonance in the 1.4 \MGEV\ region corresponding to the
$\pi(1400)$ state observed
previously\cite{thompson,long_paper,cbarrel}. However, since the
fits are satisfactory without the $\pi(1400)$, its production in
this final state is not required in the analysis.)

Many acceptable mass-dependent fits ($\chi^2/dof<1.5$) were
obtained for each of the three fit types.  The resonance
parameters from these fits were all consistent with each other.
Results from all of these fits were thus retained to determine
resonance parameters. The resonance parameters are given in
Table~\ref{mdf_par} for the $P_+$ exotic resonance and for the
lower-mass resonances in the $D_+$ and $G_+$ waves. In this table,
the central value for the mass and width of each resonance as well
as the statistical error in these quantities were determined as
the average of those quantities over all acceptable fits.
The first error in these values is statistical, determined using
the covariance matrix of the mass-independent PWA; the second is
systematic. The systematic errors are based on the range of values
allowed by taking into account different assumptions for the
partial widths of the states, different parameterizations of the
$D_+$ wave, and different ambiguous solutions.  The experimental
resolution has not been unfolded.

The mass and the width of the $P_+$ state  are consistent with
those of the $\pi_1(1600)$ exotic state observed in the
$\pi^+\pi^-\pi^-$ system\cite{sasha}.  Our data are thus
consistent with the observation of a second decay mode of the
$\pi_1(1600)$.

The first pole in the $D_+$ partial wave has mass and width
consistent with those of the $a_2(1320)$. The $D_+$ wave in the
mass region above 1.4 \MGEV\ is consistent with a broad
Breit-Wigner function centered around 1.8 -- 1.9 \MGEV\ and with
width between 0.55 and 0.75 \MGEV\ . Alternatively, that mass
region can  be described with two narrower Breit-Wigner functions.
The alternative parameterizations of the $D_+$ wave do not affect
the conclusions of this paper.

The $G_+$ partial wave has been parameterized with two
Breit-Wigner functions in fit types 2 and 3.
The parameters of the lower-mass state given in
Table~\ref{mdf_par} are consistent with the mass and width of the
$a_4(2040)$\cite{book:pdb}, which has  not been observed
previously in the $\eta^\prime\pi^-$ system. The second
Breit-Wigner pole in the $G_+$ wave is located in the high-mass
region ($\approx 2.4$ \MGEV)~ where  limited statistics results in
sizable statistical and systematic uncertainty.  Its physical
interpretation is unclear.

\begin{center}
\begin{figure}
    \psfig{figure=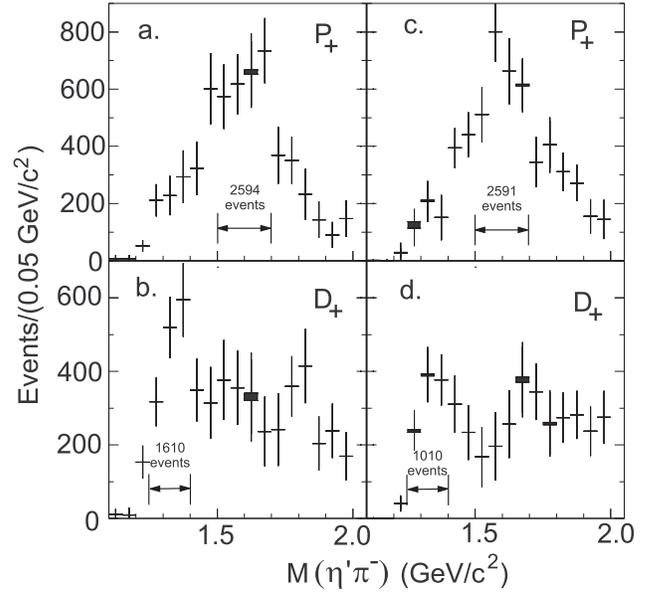,width=8.6cm,angle=0}
    \caption{(a) and (b): Mass-independent PWA for
$0.090 < |t| < 0.293~{\rm GeV}^2/c^2$;
(c) and (d): Mass-independent PWA for
 $0.293<|t|< 2.5~{\rm GeV}^2/c^2$.  The range of the ambiguous
solutions is plotted with black rectangles. \label{fit_t}}
\end{figure}
\end{center}

Mass-independent analyses have also been performed for two
separate four-momentum-transfer intervals including equal numbers
of events (see Fig.~\ref{fit_t}). The fits show that, as $|t|$
increases, the production rate for the $a_2(1320)$ decreases
faster than the production rate for the exotic state.  Note from
Fig.~\ref{fit_t} that the numbers of events in the marked peak
bins for $P_+$ production are nearly equal for the two $|t|$
intervals (the ratio of the high-$|t|$ to the low-$|t|$ peaks is
1.00 $\pm$ 0.12) whereas for $D_+$ production, the ratio is 0.71
$\pm$ 0.15. Since the $|t|$ distribution is correlated with the
production mechanism for peripheral processes, we conclude that
 exotic meson production proceeds via a different production
mechanism than that for production of the $q\overline{q}$
$a_2(1320)$ meson, or that it proceeds with a different mixture of
the same production mechanisms.

In conclusion, we have studied the $\eta^\prime\pi^-$ system
produced in the reaction $\pi^-p\rightarrow p\eta^\prime\pi^-$ at
18 \GeVc\ .  We find that an exotic meson, the $\pi_1(1600)$ is
produced, decaying to $\eta^\prime\pi^-$. The different t
dependence for their production shows that the well-known
$a_2(1320)$ and the exotic $\pi_1(1600)$ are produced via
different production mechanisms. Finally, a high-mass state
consistent with the $a_4(2040)$ has been observed decaying to
$\eta^\prime\pi^-$.



We would like to express our appreciation to the members of the
MPS group,
and to the staffs of the AGS, BNL, and  the  collaborating
institutions for their  efforts. This research was supported in
part by the National Science Foundation, the US Department of
Energy, and the Russian State Committee for Science and
Technology. The Southeastern Universities Research Association
(SURA) operates the Thomas Jefferson National Accelerator Facility
for the United States Department of Energy under contract
DE-AC05-84ER40150.

\begin{table}
\caption{Fitted Resonance Parameters }
\begin{tabular}{c|c|c}
Partial Wave & Mass & Width  \\ \hline $P_+ $  &  $1.597\pm
0.010^{+0.045}_{-0.010}$ & $0.340\pm 0.040\pm 0.050$ \\ $D_+ $  &
$1.318\pm 0.008 ^{+0.003}_{-0.005}$ & $0.140\pm 0.035 \pm 0.020 $
\\ $G_+ $ &  $2.000\pm 0.040 ^{+0.060}_{-0.020}$ & $0.350\pm
0.100^{+0.070}_{-0.050}$ \\
\end{tabular}
\label{mdf_par}
\end{table}

\end{document}